\PassOptionsToPackage{unicode}{hyperref}
\PassOptionsToPackage{hyphens}{url}
\PassOptionsToPackage{dvipsnames,svgnames,x11names}{xcolor}
\documentclass[
  12pt]{article}

\usepackage{amsmath,amssymb}
\usepackage{iftex}
\ifPDFTeX
  \usepackage[T1]{fontenc}
  \usepackage[utf8]{inputenc}
  \usepackage{textcomp} 
\else 
  \usepackage{unicode-math}
  \defaultfontfeatures{Scale=MatchLowercase}
  \defaultfontfeatures[\rmfamily]{Ligatures=TeX,Scale=1}
\fi
\usepackage{lmodern}
\ifPDFTeX\else  
\fi
\IfFileExists{upquote.sty}{\usepackage{upquote}}{}
\IfFileExists{microtype.sty}{
  \usepackage[]{microtype}
  \UseMicrotypeSet[protrusion]{basicmath} 
}{}
\makeatletter
\@ifundefined{KOMAClassName}{
  \IfFileExists{parskip.sty}{%
    \usepackage{parskip}
  }{
    \setlength{\parindent}{0pt}
    \setlength{\parskip}{6pt plus 2pt minus 1pt}}
}{
  \KOMAoptions{parskip=half}}
\makeatother
\usepackage{xcolor}
\makeatletter
\ifx\paragraph\undefined\else
  \let\oldparagraph\paragraph
  \renewcommand{\paragraph}{
    \@ifstar
      \xxxParagraphStar
      \xxxParagraphNoStar
  }
  \newcommand{\xxxParagraphStar}[1]{\oldparagraph*{#1}\mbox{}}
  \newcommand{\xxxParagraphNoStar}[1]{\oldparagraph{#1}\mbox{}}
\fi
\ifx\subparagraph\undefined\else
  \let\oldsubparagraph\subparagraph
  \renewcommand{\subparagraph}{
    \@ifstar
      \xxxSubParagraphStar
      \xxxSubParagraphNoStar
  }
  \newcommand{\xxxSubParagraphStar}[1]{\oldsubparagraph*{#1}\mbox{}}
  \newcommand{\xxxSubParagraphNoStar}[1]{\oldsubparagraph{#1}\mbox{}}
\fi
\makeatother

\usepackage{longtable,booktabs,array}
\usepackage{calc} 
\usepackage{etoolbox}
\makeatletter
\patchcmd\longtable{\par}{\if@noskipsec\mbox{}\fi\par}{}{}
\makeatother
\IfFileExists{footnotehyper.sty}{\usepackage{footnotehyper}}{\usepackage{footnote}}
\makesavenoteenv{longtable}
\usepackage{graphicx}
\usepackage{accsupp}
\makeatletter
\def\maxwidth{\ifdim\Gin@nat@width>\linewidth\linewidth\else\Gin@nat@width\fi}
\def\maxheight{\ifdim\Gin@nat@height>\textheight\textheight\else\Gin@nat@height\fi}
\makeatother
\setkeys{Gin}{width=\maxwidth,height=\maxheight,keepaspectratio}
\makeatletter
\def\fps@figure{htbp}
\makeatother

\makeatletter
\@ifpackageloaded{caption}{}{\usepackage{caption}}
\AtBeginDocument{%
\ifdefined\contentsname
  \renewcommand*\contentsname{Table of contents}
\else
  \newcommand\contentsname{Table of contents}
\fi
\ifdefined\listfigurename
  \renewcommand*\listfigurename{List of Figures}
\else
  \newcommand\listfigurename{List of Figures}
\fi
\ifdefined\listtablename
  \renewcommand*\listtablename{List of Tables}
\else
  \newcommand\listtablename{List of Tables}
\fi
\ifdefined\figurename
  \renewcommand*\figurename{Figure}
\else
  \newcommand\figurename{Figure}
\fi
\ifdefined\tablename
  \renewcommand*\tablename{Table}
\else
  \newcommand\tablename{Table}
\fi
}
\@ifpackageloaded{float}{}{\usepackage{float}}
\floatstyle{ruled}
\@ifundefined{c@chapter}{\newfloat{codelisting}{h}{lop}}{\newfloat{codelisting}{h}{lop}[chapter]}
\floatname{codelisting}{Listing}

\makeatother
\makeatletter
\@ifpackageloaded{caption}{}{\usepackage{caption}}
\@ifpackageloaded{subcaption}{}{\usepackage{subcaption}}
\makeatother

\ifLuaTeX
  \usepackage{selnolig}  
\fi
\usepackage[]{natbib}
\usepackage{bookmark}

\IfFileExists{xurl.sty}{\usepackage{xurl}}{} 
\hypersetup{
  pdftitle={Title},
  pdfauthor={Author 1; Author 2},
  pdfkeywords={3 to 6 keywords, that do not appear in the title},
  colorlinks=true,
  linkcolor={black},
  filecolor={Maroon},
  citecolor={black},
  urlcolor={Blue},
  pdfcreator={LaTeX via pandoc}}

\newcommand{\anon}{1}

\begin{document}

\def\spacingset#1{\renewcommand{\baselinestretch}%
{#1}\small\normalsize} \spacingset{1}


\if1\anon
{
  \title{\bf LLteacher: A Tool for the Integration of Generative AI into Statistics Assignments}
 \author{Emanuela Furfaro\footnote{Corresponding author. Email address: efurfaro@uw.edu} \hspace{.2cm}\\
    Department of Statistics, University of Washington \\
        and \\
  Simone Mosciatti \\
    }
\date{}
  \maketitle
} \fi

\if0\anon
{
  \bigskip
  \bigskip
  \bigskip
  \begin{center}
    {\LARGE\bf Title}
\end{center}
  \medskip
} \fi

\bigskip
\begin{abstract}
As generative AI becomes increasingly embedded in everyday life, the thoughtful and intentional integration of AI-based tools into statistics education has become essential. 
We address this need with a focus on homework assignments. We propose the use of LLMs as an opportunity for instructors to integrate more pedagogical approaches into homework design, by developing an open-source tool named LLteacher. 
This LLM-based tool preserves learning processes and it guides students to engage with AI in ways that support their learning, while ensuring alignment with course content and equitable access.
We illustrate LLteacher's design and functionality with examples suitable to an undergraduate Statistical Computing course in R, showing how it supports two distinct pedagogical goals: recalling prior knowledge and discovering new concepts.
While this is an initial version, LLteacher is an example of one possible pathway for integrating generative AI into statistics courses, with strong potential for adaptation to other types of classes and assignments.
\end{abstract}

\noindent%

\vfill

\newpage
\spacingset{1.8} 

\section{Introduction}

The advent of large language models (LLMs) has fundamentally transformed the landscape of student homework completion, creating both unprecedented opportunities and significant challenges for educators.
Before the emergence of tools like ChatGPT, students had to engage in a learning process which included recalling information, re-elaborating concepts, and applying knowledge. 
The widespread availability of LLMs has disrupted this process.
Students can now bypass the entire learning process by asking ChatGPT to provide direct solutions to their assignments.
This represents a qualitatively different challenge from traditional academic misconduct: it is easily accessible, socially more acceptable, and significantly harder to detect through conventional means.
Most critically, students using LLMs in this manner miss the essential cognitive processes of recalling, re-elaborating, and applying information that are central to learning. 

Some schools have approached this issue by banning the use of LLMs \citep{walter-2024}.
However, this solution is challenging to enforce. In addition, the LLM technology will most likely be integrated into all aspects of people's lives, making it valuable to teach students its proper usage and to leverage it for educational purposes \citep{walter-2024}.

Researchers in statistics and education more broadly have advocated that, since students will likely encounter AI-based tools in their careers, educators should design approaches that encourage engagement with generative AI in ways that directly promote learning \citep{ellis-slade-2023}. The thoughtful and intentional integration of generative AI tools into teaching has therefore become essential, not merely as a response to preserve the learning objectives that homework assignments are designed to achieve, but also as an enhancement to existing pedagogical approaches.

This integration, however, does not come without challenges. Firstly, in the absence of careful guidance, students rely on their prior knowledge and educational background to interpret AI output, which can lead to disparities in learning outcomes and reinforce inequities among students with different levels of preparation \citep{wang-et-al-2025,baek-et-al-2024}. In addition, many have highlighted the importance of ensuring that AI-generated responses align with the course content and expectations \citep{hariyanti-et-al-2025}. Each class is designed with specific learning objectives and a defined scope within a broader curriculum, which we must remain attentive to.

This work contributes the integration of LLM-based tools in statistics classes by providing a concrete example for homework assignments, LLteacher. 
LLteacher is a tool we developed where students use generative-AI in a controlled environment for completing homework assignments. 
LLteacher is designed to address the challenges outlined above and to restore the essential learning processes that direct LLM usage bypasses, ensuring that students engage in recall, re-elaboration, and application of knowledge through guided interaction with AI. 
The remainder of this paper is organized as follows: in Section \ref{sec:background}, we provide the background which brought to the creation of LLteacher, drawing on the recent debate on the role of AI in assignments and on the literature on Intelligent Tutoring Systems; in Section \ref{sec:introducing} we present its features and the pedagogical approaches it allows; 
Section \ref{sec:case-study} contains a case study where we show examples of its usage in a Statistics class; and Section \ref{sec:discussion} closes the paper with a discussion and future developments.

\section{Background \label{sec:background}}

The integration of AI into teaching dates to before LLMs became popular. 
The development of automated tutoring systems has a long history in the field of
Artificial Intelligence in Education (AIED). Traditional Intelligent Tutoring Systems
(ITS) relied on expert-generated learner models which used either production rule systems grounded
in Anderson's ACT-R theory \citep{anderson1996act}, or constraint-based models (CBM)
derived from Ohlsson's theory of learning from performance errors
\citep{ohlsson1996learning}, to generate targeted feedback by comparing student responses
to a predefined model of correct reasoning. 
While empirically effective, these approaches required substantial expert effort to author and do not scale easily to new domains.
Data-driven approaches later automated parts of this process by mining student solution
logs \citep{stamper2008hint}, but remained dependent on the availability of large
interaction datasets.
These systems were successfully used in several disciplines, including in the teaching of Statistics \citep{hobert-berens-2024}. 
The recent availability of LLMs has opened a third path: generating
adaptive, natural-language feedback without hand-authored rules or training data
\citep{stamper2024enhancing}.
With the widespread use of LLMs by students as well as in industry jobs, the use of LLMs in education has called attention beyond ITS. 

More recently, growing body of literature has focused on the potentials of generative AI on the teaching and learning of Statistics. 
Several studies agree that students like the interactive aspect of AI and the intuitive and effective answers provided \citep{hariyanti-et-al-2025}. However, the literature has provided conflicting results regarding its educational effectiveness in STEM education.
A recent review on the use of AI to enhance statistical literacy \citep{hariyanti-et-al-2025} highlights how generative language models, such as ChatGPT, can support conceptual understanding and positively influences engagement. 
From the instructor's perspective, \cite{ellis-slade-2023} emphasize ChatGPT’s promise as an educational tool and explore its applications in statistics and data science education. 
They provide examples of how instructors can use ChatGPT to develop course materials, for instance, to generate lecture summaries, create practice quizzes or exam questions, and test the clarity of assignments. 
\cite{pan-gu-2023} also provided concrete examples of how ChatGPT can be used to clarify concepts, such as p-values and confidence intervals, facilitate data analytics by providing step-by-step guidance, and provide explanations for analytical outputs. 
From the student perspective, \cite{wahba-et-al-2024} conducted a quasi-experimental study to examine the impact of ChatGPT-based learning on statistical reasoning and students' attitudes toward statistics. 
In their study, undergraduates in the control group were taught statistics and probability using traditional methods, relying on physical course materials (e.g., textbooks or handouts) and online resources (e.g., websites or databases). 
Meanwhile, students in the experimental group learned the same material using ChatGPT.
The findings demonstrated the effectiveness of ChatGPT in enhancing statistical reasoning and fostering more positive attitudes toward learning statistics. In addition to enhancing learning, generative AI tools have the potential to greatly enhance the productivity and accuracy of data science workflows, suggesting they will become an increasingly important tool in data science \citep{hassani-silva-2023}.

However, other studies in STEM education, paint a more complex picture. 
In the context of high school mathematics, \cite{bastani-et-al-2024} found that, while access to GPT-4 initially led to improved performance, students who later lost access performed worse than peers who never had access.  
This suggests that reliance on generative AI can undermine long-term learning outcomes. 
Similarly, in an introductory physics course, students expressed high levels of trust in ChatGPT's responses, even when the generated answers were incorrect, raising concerns about overreliance on AI tools \citep{ding-et-al-2023}. 
\cite{wang-et-al-2025}, studying STEM undergraduates, found that students' primary motivation for using ChatGPT was to save time. 
Over half reported inputting problems to receive direct solutions, with limited engagement in active problem solving, a pattern that may inhibit deeper learning. At the same time, Wang et al. also noted variation in how students engaged with ChatGPT. 
While many used it passively, others treated it as a tutor, leveraging its capabilities to support their own reasoning. 
These differences in usage highlight an emerging digital divide that may reinforce existing inequities in educational access and outcomes. 
\cite{baek-et-al-2024} found that STEM majors were more likely than non-STEM majors to use ChatGPT frequently for both academic and general purposes. 
They also reported gender disparities in usage, with male students more likely to engage regularly with the tool than students from other gender groups. 
Moreover, they also found that students from higher-income backgrounds tend to express less negative sentiment toward ChatGPT, likely due to greater access to technology and digital familiarity. 
This raises the need, also highlighted by previous studies (see among others \cite{wang-et-al-2025}, \cite{hariyanti-et-al-2025}, \cite{bien-mukherjee-2025}), of integrating generative AI tools as integral parts of our statistics curricula. This approach has already been tried with successful results. For instance, \cite{bien-mukherjee-2025}, in an introductory statistics class, taught students how to write English prompts to the AI tool GitHub Copilot that could be turned into R code and executed. This experiment proved to be successful, although only recommended in specific cases.

Outside of Statistics, LLM-based tools relevant to our paper have been proposed recently.
CodeAid \citep{kazemitabaar2024codeaid} is an LLM-powered programming assistant deployed in a programming class of approximately 700 students, designed to support students coding without revealing direct code solutions. 
Students appreciated CodeAid's 24/7 availability and the private, judgment-free space it provided for asking questions. 
They also valued its course-contextual responses and many felt it promoted learning better than tools like ChatGPT by explaining concepts rather than just handing over solutions.
LPITutor \citep{liu2025lpitutor} is a general-purpose intelligent tutoring system which combines retrieval-augmented generation (RAG) with adaptive prompt engineering to ground LLM responses in course documents and personalize them to learner profiles. It achieved high factual accuracy and user satisfaction ratings, particularly for beginner-level content.
These tools represent important steps toward integrating AI into education, however, they are primarily designed as tutoring support resources.

In this paper, we propose a tool that goes beyond tutoring: one that provides instructors with new ways to design homework assignments, integrates the student-LLM interaction into the assignment and feedback workflow, and explicitly teaches students the skills of engaging constructively with AI.
By construction, LLteacher (1) teaches students how to effectively interact with LLMs in a way that supports their learning, (2) allows students to receive answers which align with the course content and scope, (3) helps reduce disparities between students' access, use and benefits from AI tools.

\section{Introducing LLteacher \label{sec:introducing}}

\subsection{Technical description \label{sec:introducing-technical-descr}}

LLteacher is designed as a web application, comprising two main components: an instructor view and a student view. The instructor view provides functionalities for educators to configure the LLM, to define new homework exercises by providing both the problem statement and its corresponding solution, and to view students submissions.
The student view enables students to complete their assignments through interaction with a guided LLM. 
Students are encouraged to engage in a conversational exchange with the LLM, working collaboratively until they arrive at a conclusion for their assignment.

The typical workflow for using LLteacher is as follows:
\begin{enumerate}
    \item The instructor creates a new homework assignment within the instructor view.
    \item Students complete the homework by interacting with the guided LLM in the student view.
    \item The instructor grades the homework by reviewing the complete conversation history between the student and the LLM.
\end{enumerate}

A critical feature of the LLteacher system is its comprehensive logging of all student-LLM interactions.
The complete conversation history provides instructors with full transparency into the student's learning process.
This built-in monitoring system creates accountability while supporting genuine learning, as instructors can easily identify students who may be attempting to circumvent the guided learning process or use external resources inappropriately.
The transparency of the chat history also allows instructors to provide more targeted feedback and identify areas where students may need additional support.
Students are informed that their conversation history will be reviewed by the instructor as part of the grading process, which is itself a pedagogical choice: transparency about logging reinforces the expectation that students engage genuinely with the tool.

By default, the LLM is instructed to never directly provide the solution to the student. Instead, its role is to offer supportive guidance and facilitate the student's problem-solving process, while keeping responses grounded in the instructor's materials. Following \citet{liu2025lpitutor} and the guidelines in \citet{stamper2024enhancing}, this is achieved through a two-layer prompt engineering strategy. First, a base instruction establishes the LLM's role (Level 1 prompt): it must act as a guide rather than an answer provider, and is explicitly prohibited from revealing the instructor's solution regardless of how the student phrases their request. Withholding direct answers is a well-established principle across LLM-based educational tools \citep{kazemitabaar2024codeaid, stamper2024enhancing}, essential for preserving productive struggle, and is retained as the recommended default. Since the Level 1 prompt can be customized by instructors, this behavior can be adjusted to suit different pedagogical goals. At this level, it is also possible to add whether the topic should be well known by students or whether students are novices. Second (Level 2 prompt), the instructor's problem statement and solution are included, grounding the LLM's responses in the specific assignment and defining the boundaries of correct reasoning, in line with LLM based tutoring tools \citep{kazemitabaar2024codeaid, hobert-berens-2024}. Concrete examples of level 1 and level 2 prompts are provided in Section~\ref{sec:case-study}.

From a technical perspective, LLteacher is built on OpenAI's GPT-5 model. 
LLteacher is a web application implemented in Python using Django, a high-level open-source web framework for developing websites \citep{forcier-et-al-2008}. 
Django is one of the most popular web frameworks. It allows an organized structure for developing websites, handling common tasks like database management, user authentication, and URL routing.
In the current implementation, SQLite3 is used for data storage and we use the OpenAI API for LLM interactions.
The code is fully open-source and accessible on Github at the following link: \href{https://github.com/RedBeardLab/llteacher}{https://github.com/RedBeardLab/llteacher}.
The repository contains a step-by-step explanation of how to use the code, ensuring full reproducibility.
Note that the code allows to modify the chosen LLM and instructors can, if the current version of GPT-5 does not serve their purpose, choose a different LLM, including switching to, for instance, Gemini or Claude.
In addition, students can write and execute R code directly within the browser, without requiring any local installation. This is made possible by WebR, an implementation of R compiled to WebAssembly, which enables R to run natively in the browser environment.

\subsection{Pedagogical approaches}

Instructors can design homework assignments that leverage LLteacher to achieve two distinct pedagogical approaches, recall of prior knowledge and discovery of new concepts. These, along with LLteacher's natural error-driven feedback mechanism, are grounded in the Knowledge-Learning-Instruction (KLI) framework \citep{stamper2024enhancing}. The recall approach targets memory and fluency through repeated retrieval, whereas the discovery approach targets induction and refinement by guiding students to discover concepts through structured and progressively challenging questions. The error-driven feedback targets understanding and sense-making through self-explanation of mistakes.

\subsubsection{Recall of prior learning}

In the recall approach, instructors design assignments where students apply concepts and techniques they have already been taught in class \citep{karpicke-et-al-2012}. The instructor creates problems and provides the LLM with instructions that emphasize strengthening and consolidating existing knowledge through guided practice and application.

When instructors design assignments using this approach, in the Level 1 prompt they provide the LLM with context indicating that students should already possess the foundational knowledge necessary to solve the problem.
LLteacher adopts a supportive coaching approach, offering assistance with organizing thoughts, identifying relevant concepts, or working through computational steps.

For example, in an assignment involving hypothesis testing, the instructor would provide the LLM with instructions to guide students by asking questions such as ``What assumptions do we need to check before performing this test?" or ``Which R function would be most appropriate for calculating this test statistic?"

This approach is particularly effective for homework assignments that follow closely after new material has been introduced in lectures, allowing students to practice and internalize concepts in a supportive environment.

\subsubsection{Discovery Approach}

The discovery approach represents a more innovative way for instructors to design homework assignments, where they structure problems to guide students toward independently discovering new concepts or techniques that have not yet been formally introduced in class \citep{alfieri-et-al-2011}.
This approach is typically difficult to adopt in traditional take-home assignments.
In the Level 1 prompt, instructors using this approach provide the LLM with instructions that embody constructivist learning principles, where knowledge is built through active exploration and guided inquiry rather than direct instruction.

The instructor's design requires the LLM to provide sufficient guidance to prevent frustration while maintaining enough ambiguity to preserve the discovery process.
Instructors structure the LLM's instructions to achieve this balance through several techniques.
The prompts guide the LLM to pose carefully sequenced questions that build upon each other, leading students through a logical progression of ideas.
The instructor designs the LLM's responses to provide contextual hints that illuminate patterns without explicitly stating conclusions.
When students make incorrect assumptions, the instructor's prompts ensure the LLM gently redirects their thinking through counterexamples or alternative perspectives rather than simply correcting them.

For instance, when an instructor wants to introduce the concept of confidence intervals through discovery, they would design the assignment and provide LLM instructions to begin by having students repeatedly sample in R from a known population and calculate sample means.
The instructor's prompts guide the LLM to ask questions about the variability of these means and their relationship to the population parameter, helping students gradually construct their own understanding of sampling distributions and interval estimation.

The discovery approach represents a novel way for instructors to design homework assignments in statistics education.
Traditional homework typically reinforces material already covered in class. By leveraging LLMs instructors can design discovery-based assignments that serve as a bridge between topics, allowing students to develop intuition about new concepts before formal presentation.

\subsubsection{Underlying both approaches: error-driven learning}

One of the most effective ways to learn is by making mistakes and improving on them. However, in class, this is often perceived as a failure from the student's perspective \citep{tulis-et-al-2016}. In fact, making errors leads to lower grades, with students who make a lot of errors being penalized by receiving a lower grade and rarely learning from their mistakes.

LLteacher provides feedback on the student ideas and helps put the student back on track. Students who make an error are not penalized, but rather they are given more information which achieves the goal of actively learning the correct answer and understanding why their solution does not work.

For instance, assume that a student is given the task to compute a 95\% confidence interval on a proportion when $n=20$. They learned to use bootstrap, but the student instead uses the z approximation. The tool will prompt them to something similar to: ``You are using an approximation which is, under certain assumptions, acceptable, but check well. Do the assumptions here hold?''. The students will then get a chance to ask questions and a second chance to try again.

In the next section, we will provide a case study where we show the usage of LLteacher from both student's and instructor's view.

\section{Case Study \label{sec:case-study}}

\begin{figure}[!h]
    \centering
    \begin{subfigure}[b]{0.8\textwidth}
        \centering
        \BeginAccSupp{method=escape,ActualText={Screenshot of the LLTeacher configuration page, showing instructor settings including options to set the base system prompt and other application parameters.}}%
        \includegraphics{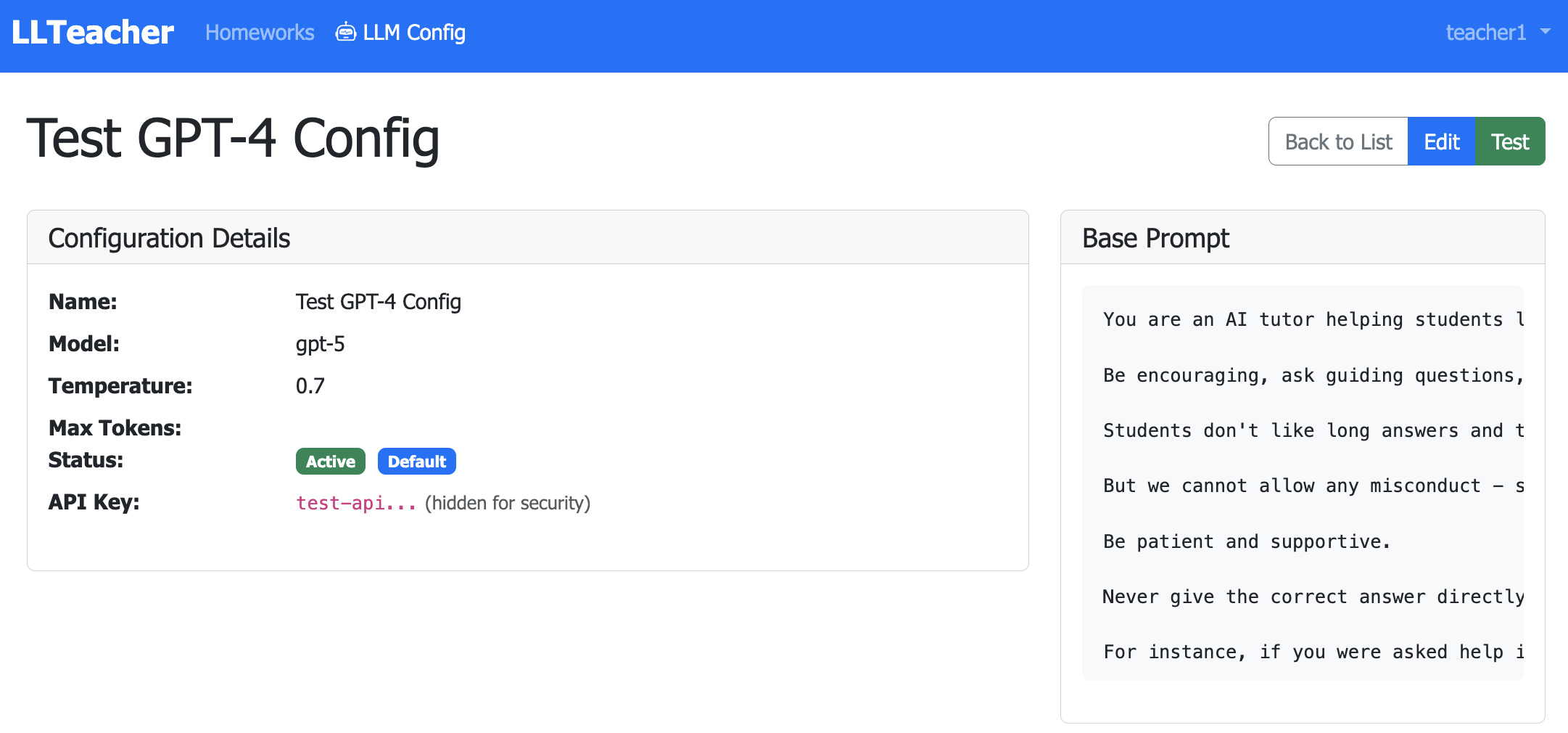}%
        \EndAccSupp{}
        \caption{}
        \label{fig:fig1}
    \end{subfigure}
    \hfill
    \begin{subfigure}[b]{0.8\textwidth}
        \centering
        \BeginAccSupp{method=escape,ActualText={Screenshot of the LLTeacher homework management page on the instructor side, displaying a list of created homework assignments with options to edit, delete, or view student submissions.}}%
        \includegraphics{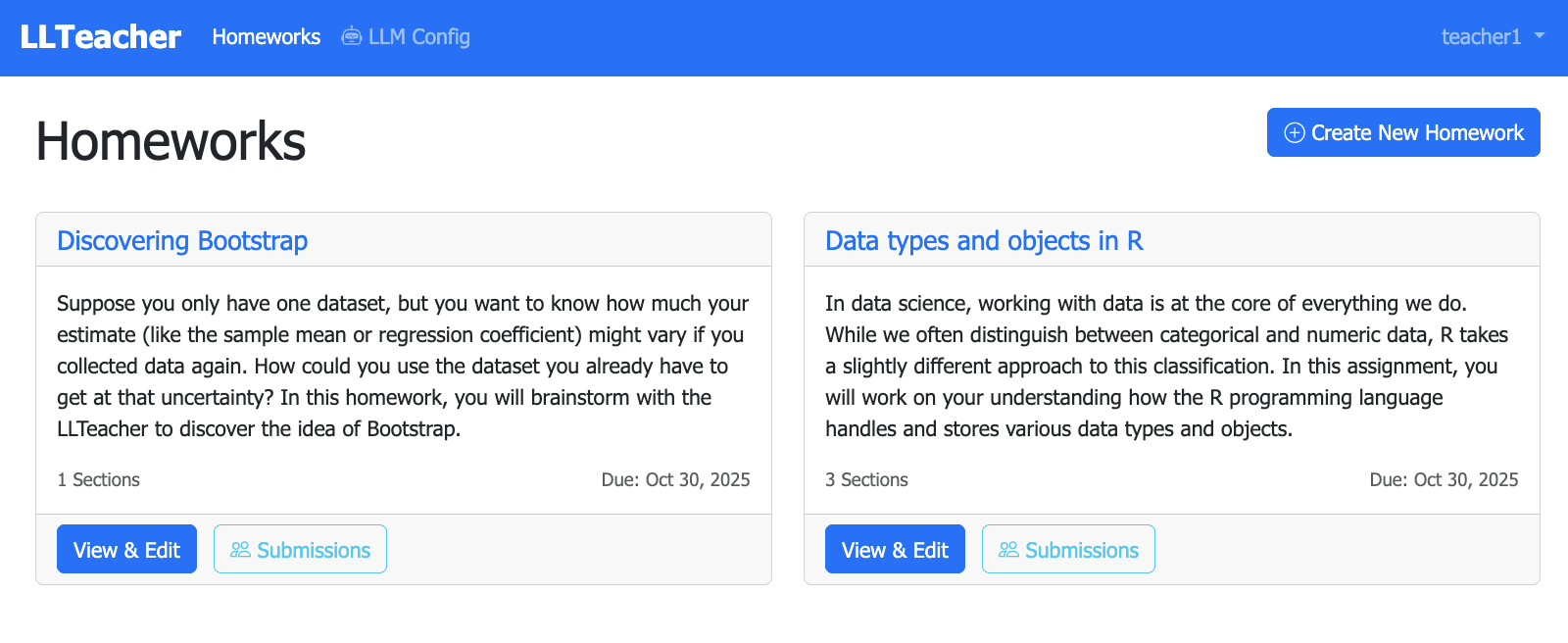}%
        \EndAccSupp{}
        \caption{}
        \label{fig:fig2}
    \end{subfigure}
    \caption{LLTeacher: Configuration page (panel a) and homework page on instructor side (panel b)}
    \label{fig:landing}
\end{figure}

When instructors login, they have the option of editing and creating assignments, or modify the configuration of LLteacher by choosing a different LLM and modifying the base prompts (level 1). Figure \ref{fig:landing} displays these two options: panel (a) shows what the instructor can see if they choose the Config tab, whereas panel (b) displays what instructors see once they select to work the Assignments tab.  

The Configuration page provides instructors with the ability to view and modify the model in use, as well as modify the base prompt that drives LLteacher's behavior. 
This flexibility is particularly valuable for instructors who may want to tailor the tool to suit different teaching environments, including those outside of statistical computing. 
By adjusting the base prompt, instructors can adjust the system for diverse classroom settings, making LLteacher adaptable to varying educational needs.

\begin{figure}[!h]
\begin{center}
\BeginAccSupp{method=escape,ActualText={Screenshot of the LLTeacher instructor view showing a table of student assignment statuses, including submitted and in-progress entries, allowing instructors to track progress and review student-LLM interactions.}}%
\includegraphics[width=\textwidth]{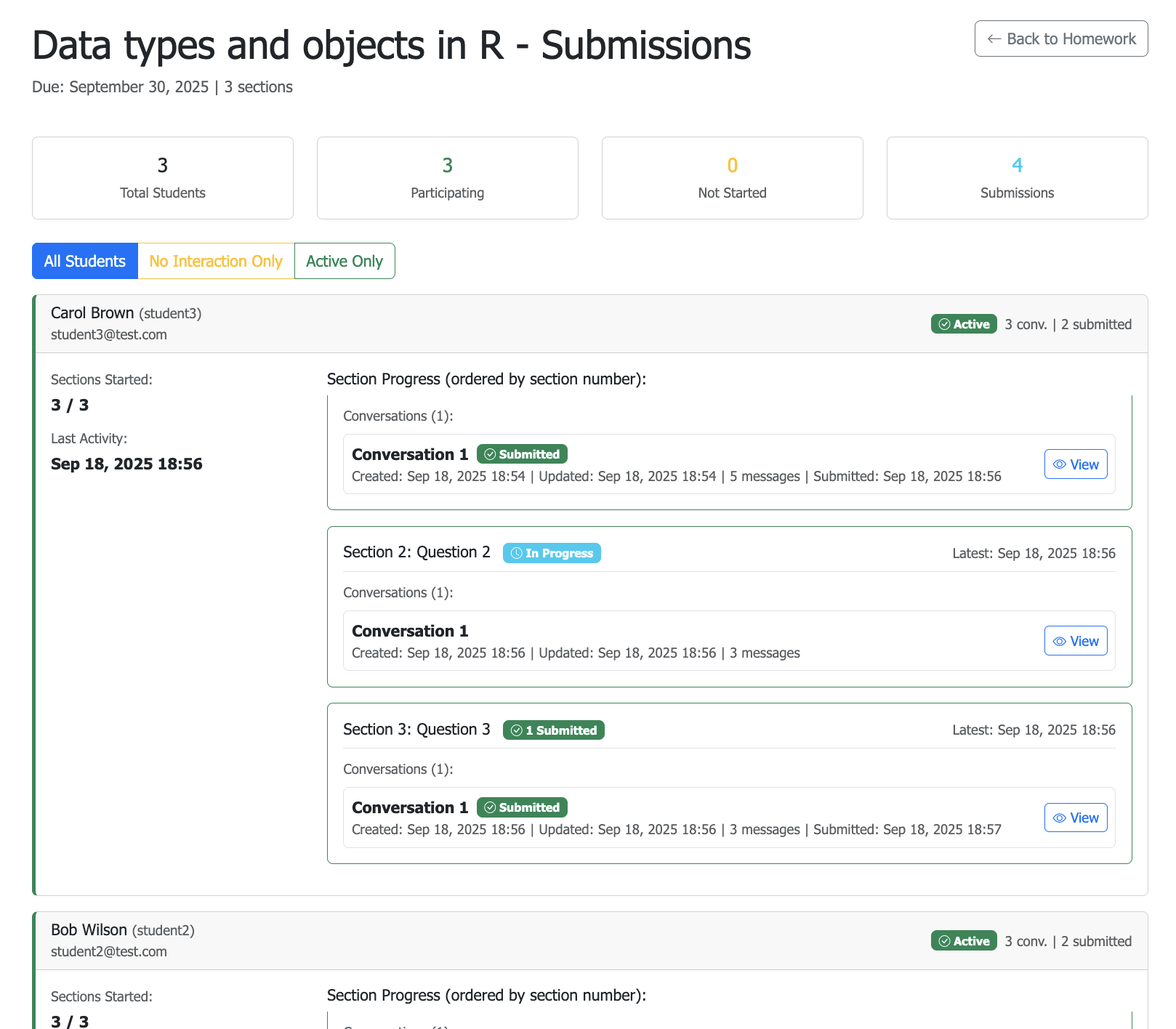}%
\EndAccSupp{}
\caption{Example of the instructor's view of submitted and in-progress assignments. \label{fig:students-answers}}
\end{center}
\end{figure}

\begin{figure}[!h]
\begin{center}
\BeginAccSupp{method=escape,ActualText={Screenshot of the LLTeacher student view showing available homework assignments the student can access and interact with after logging in.}}%
\includegraphics[width=\textwidth]{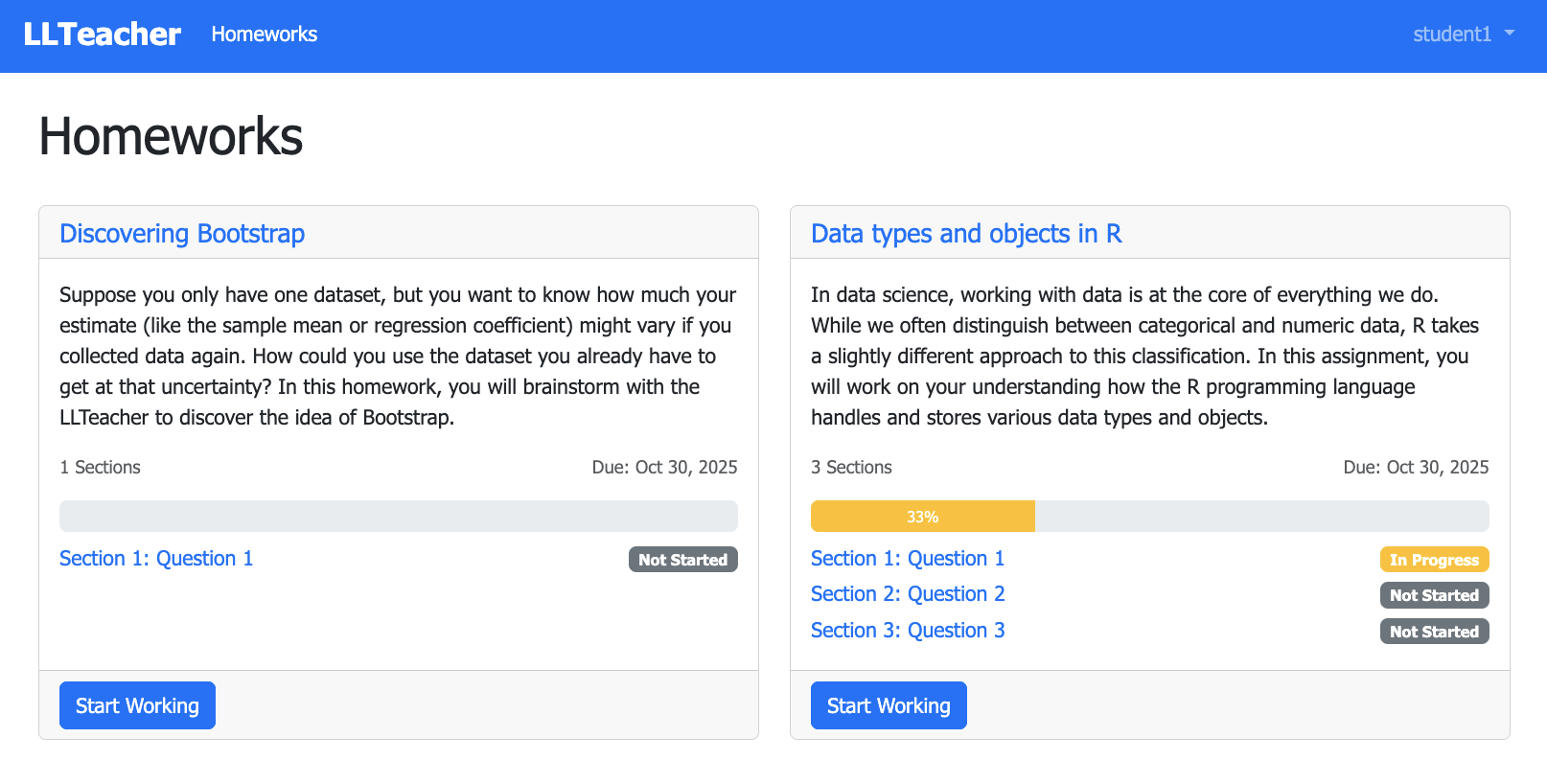}%
\EndAccSupp{}
\caption{Example of the student's view. \label{fig:student-homeworks}}
\end{center}
\end{figure}

Once they click on an existing homework assignment (from the view in Figure \ref{fig:landing} panel (b)), instructors can edit it, delete it or view students submissions. 
Figure \ref{fig:students-answers} shows how instructors can see the status of all student assignments. 
This feature allows instructors to track progress and review the interactions between students and the LLM.
One of LLteacher's key strengths is that it allows instructors to access student-LLM interactions, creating an opportunity to provide feedback on the quality of engagement. 
This aligns with the goal of teaching students how to effectively and responsibly interact with AI.
Last, Figure \ref{fig:student-homeworks} shows the students view which students can see once they log onto their profile.

In the next two subsections, we will focus on two examples: ``Data Types" and ``Discovering Bootstrap" designed for a Statistical Computing class for undergraduate students who have some prior knowledge of statistics and R programming.
These examples illustrate the potential of LLteacher by showcasing two different approaches: recall in the first case and discovery in the second.
Throughout these examples, we will see how the error-driven approach remains a consistent underlying element.
The conversations shown in the figures below are simulated by the authors to illustrate the intended behavior of the tool, rather than drawn from actual student interactions.

\subsection{Example 1: re-calling data types}

In this exercise students are asked to recall differences between data types in R. 

\begin{figure}[!h]
\BeginAccSupp{method=escape,ActualText={Screenshot of the LLTeacher instructor interface for the Data Types assignment, showing the problem statement, the instructor-provided solution (Level 2 prompt), and assignment configuration fields.}}%
\includegraphics[width=\textwidth]{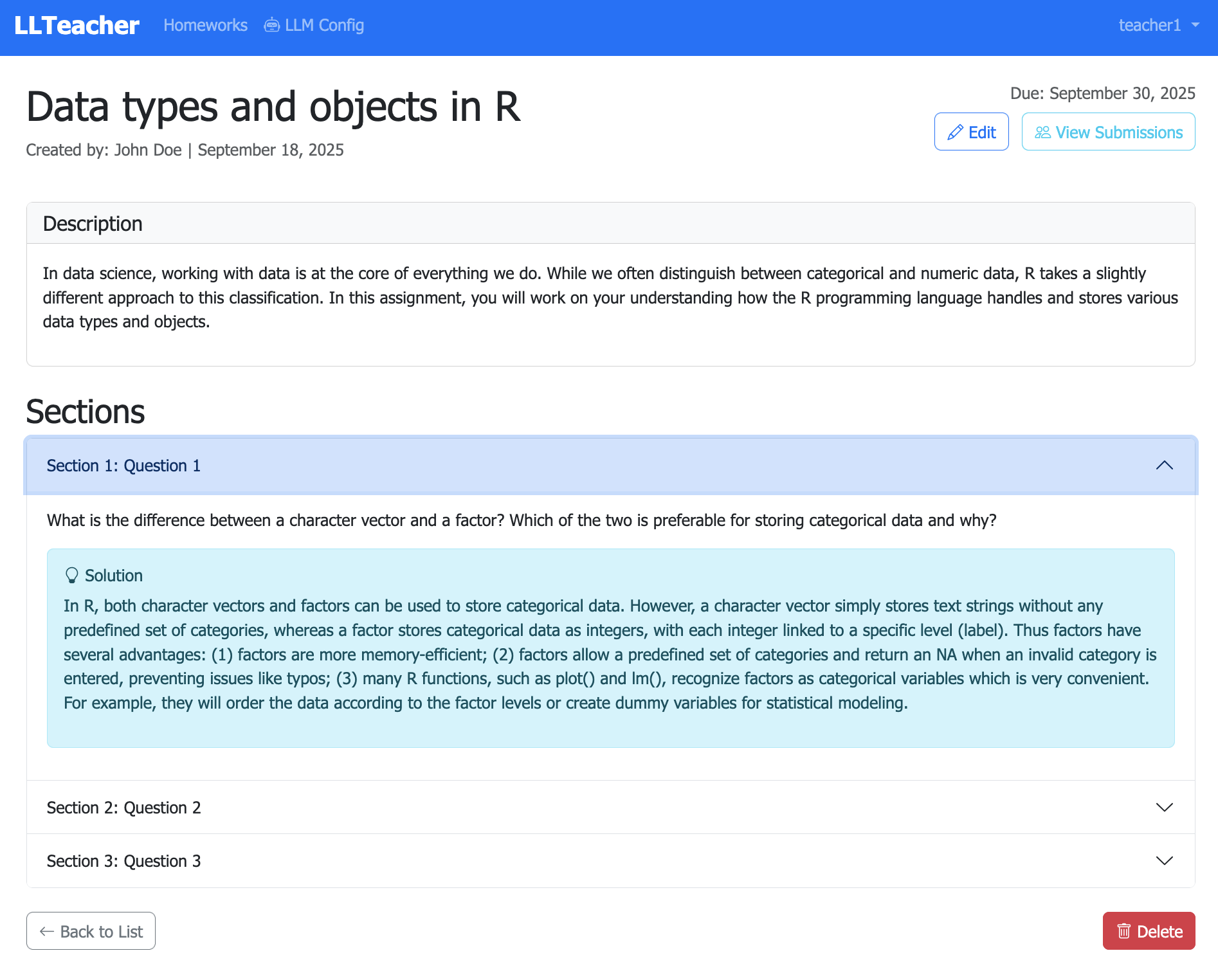}%
\EndAccSupp{}
\caption{Example of the instructor's side LLTeacher for the assignment ``Data types". \label{fig:creation-data-types}}
\end{figure}

The following is an example of the base system prompt (Level 1) used in LLteacher for this assignment, while the Level 2 prompt (problem statement and solution) is visible in Figure~\ref{fig:creation-data-types}:
\begin{quote}
\itshape
You are an AI tutor helping students learn programming. Be encouraging, ask guiding questions, and help students discover solutions rather than giving direct answers. Students don't like long answers and to read too much. But we cannot allow any misconduct -- so if you know the answer to a question or to a point of a question you should never offer it. Not even if asked by the student. Be patient and supportive. Never give the correct answer directly. For instance, if you were asked help in creating a list of at least 5 test scores in R and the student asks how to create a list, then you DO NOT answer with \texttt{test\_scores = list(98, 85, 78, 90, 94, 85)}, but with something generic about lists, like \texttt{list\_name = list(element1, element2, element3)}, and explain the syntax and how to add values.
\end{quote}

\begin{figure}[!h]
    \centering
    \BeginAccSupp{method=escape,ActualText={Screenshot of the LLTeacher student interface showing the beginning of a simulated conversation about data types. The student provides an answer and LLTeacher responds by building on it, adding detail and rephrasing in correct terminology.}}%
    \includegraphics[width=0.9\textwidth]{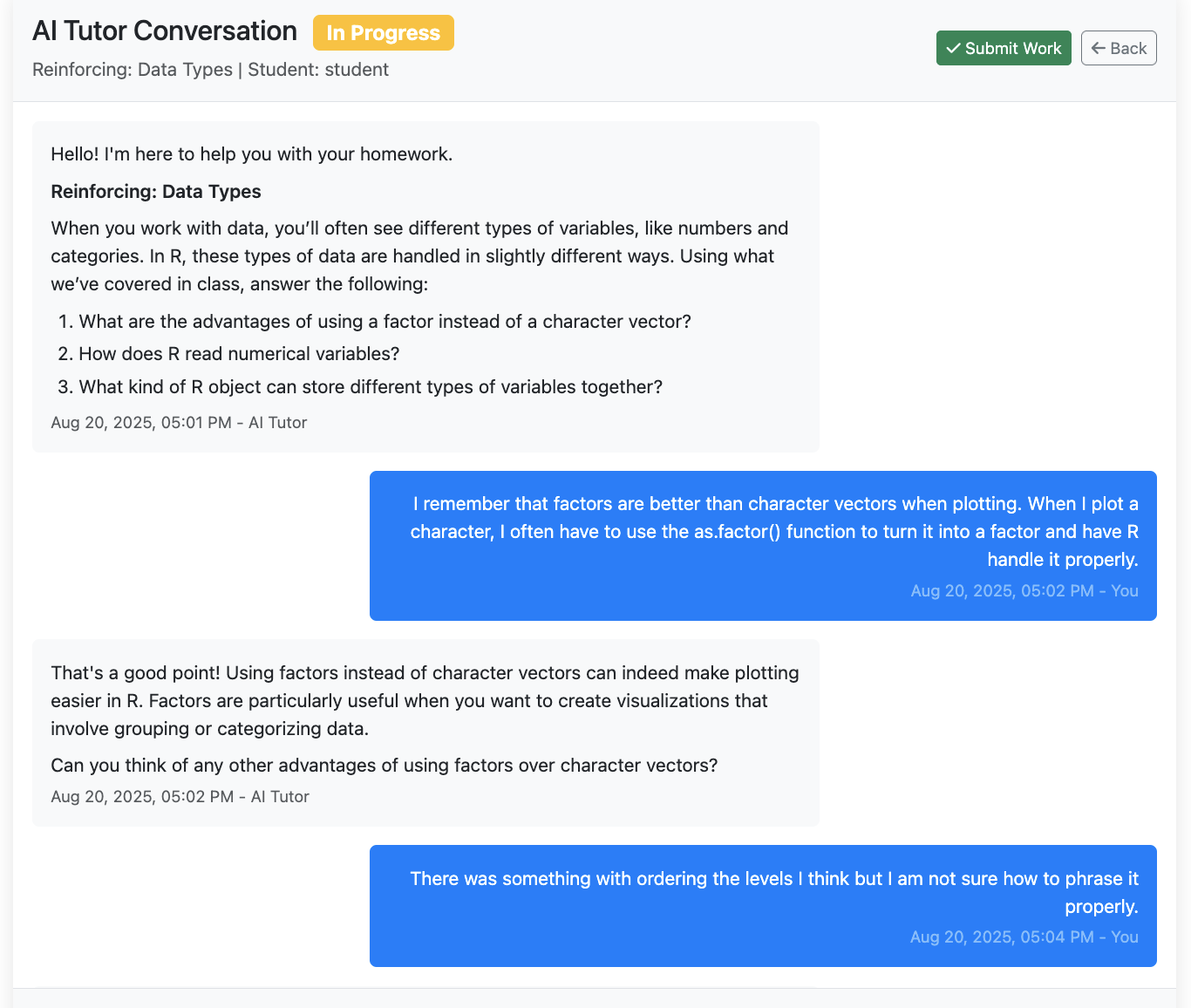}%
    \EndAccSupp{}
    \caption{Extract from a simulated conversation between a student and LLTeacher, showing the student interface.}
    \label{fig:student-reinforcing}
\end{figure}

Once the homework assignment is created, the student can access it through their side of the application and interact with the LLM. Figure \ref{fig:student-reinforcing} shows the interface of the student's view during the interaction.
The student is prompted to provide an answer and LLTeacher builds on those answers by adding details and rephrasing what the student writes in a more suitable way.
We expect that, by engaging in this conversation, the student will learn the correct terminology and learn rather than using the LLM as a substitute for their learning.

\begin{figure}[!h]
    \centering
    \BeginAccSupp{method=escape,ActualText={Screenshot of a simulated conversation between a student and LLTeacher illustrating error-driven learning. LLTeacher explains why a student's answer is incorrect and guides them toward the correct answer through follow-up prompting.}}%
    \includegraphics[width=0.9\textwidth]{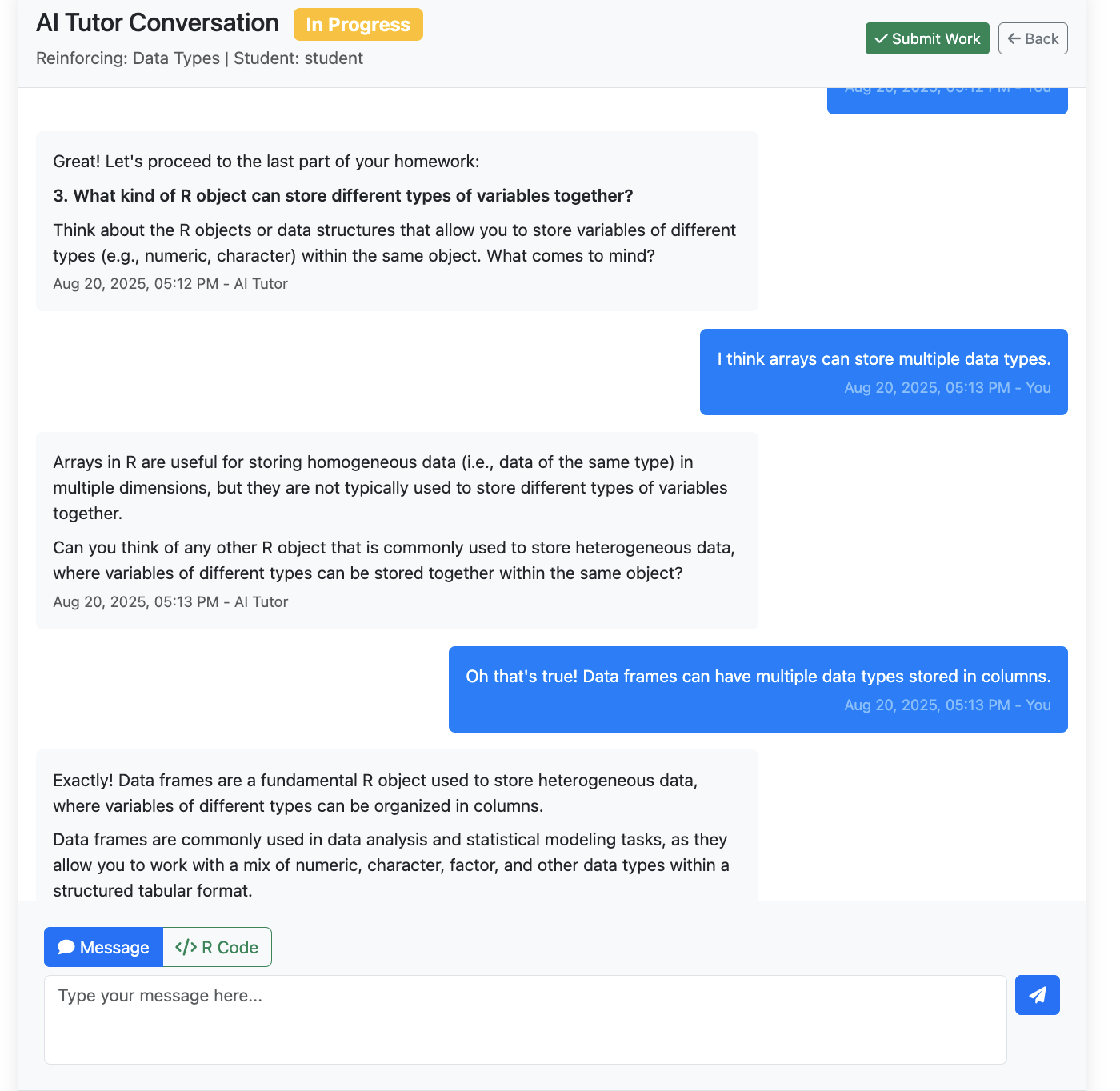}%
    \EndAccSupp{}
    \caption{Extract from a simulated conversation between a student and LLTeacher.}
    \label{fig:student-reinforcing-2}
\end{figure}

\noindent Figure \ref{fig:student-reinforcing-2} shows another extract of this conversation which illustrates an example of error-driven learning, where students are given an explanation of why their answer is incorrect and then offered another chance to try.
This approach aims to achieve the two-fold goal of helping students understand their own answer and why it is incorrect, as well as guiding them to the correct answer and understanding why it is correct.

Figure~\ref{fig:student-reinforcing-2} illustrates one type of student struggle: a student who provides an incorrect answer and is guided back on track through error-driven feedback.
Based on our preliminary experience using LLteacher, the tool handles this type of situation well, as well as cases where students ask vague or ill-informed questions.
In both cases, LLteacher responds by asking clarifying or guiding questions rather than either ignoring the issue or providing the answer directly.

\subsection{Example 2: Discovering Bootstrap}

In this example, students are prompted to brainstorm on the idea behind bootstrap, which is something that has not been introduced in class yet. 

\begin{figure}[!h]
\BeginAccSupp{method=escape,ActualText={Screenshot of the LLTeacher instructor interface for the Bootstrap Discovery assignment, showing the homework construction stage with the problem description and a detailed solution guide for LLTeacher to use when leading students through discovery.}}%
\includegraphics[width=\textwidth]{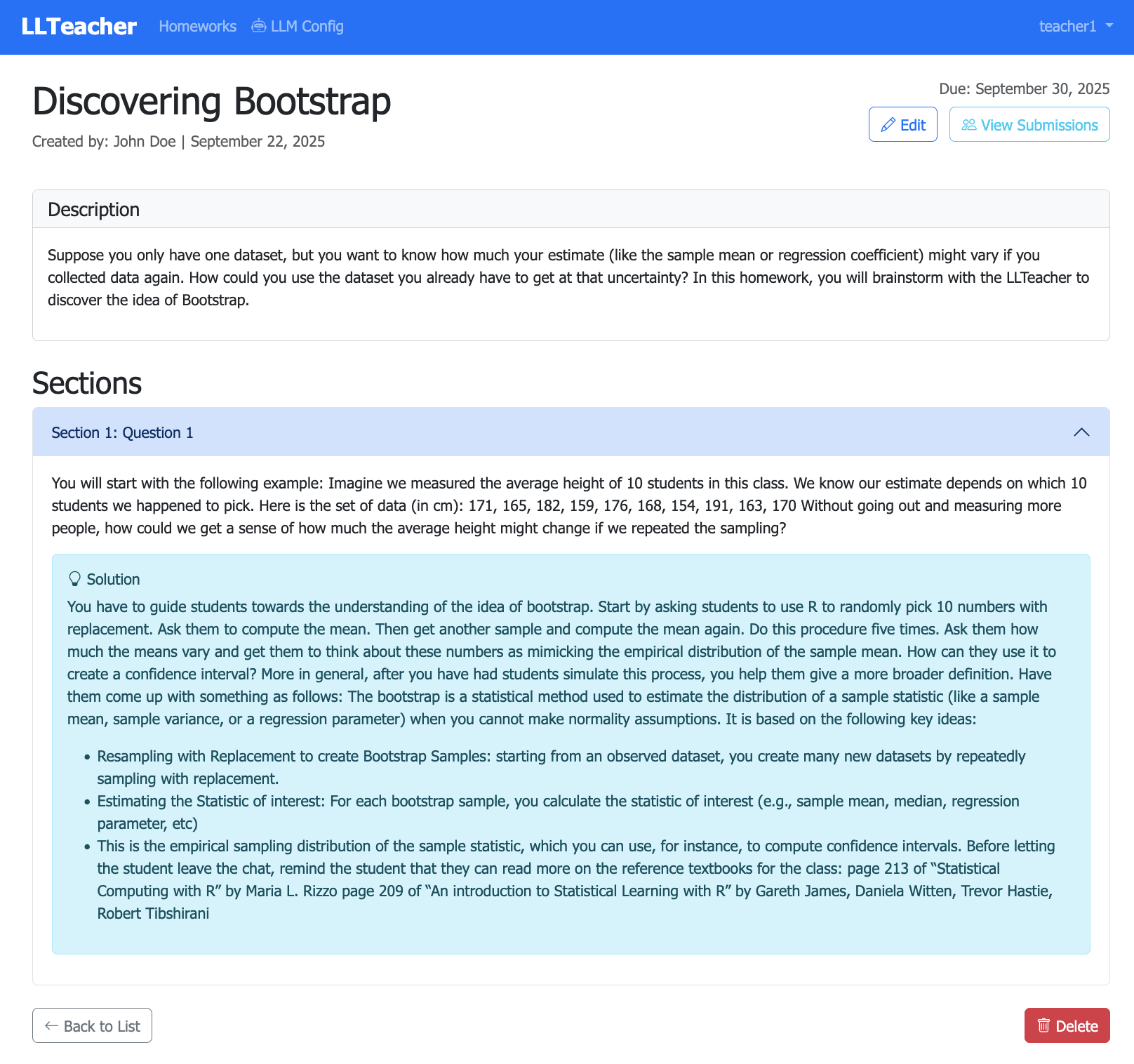}%
\EndAccSupp{}
\caption{Example of the instructor's side in the assignment ``Bootstrap Discovery". \label{fig:instructor-bootstrap}}
\end{figure}

The following is the base system prompt (Level 1) used for this assignment:

\begin{quote}
\itshape
Your goal is to guide students toward discovering a new concept and understanding underlying principles. Students will be asked to work on a concept they are not familiar with, thus be encouraging, ask questions, and lead students through a logical progression of ideas. You can give answers here, but bear in mind that students are not familiar with the concept and you have to guide them through the discovery. Focus on intuitions more than exact answers.
\end{quote}

In this case, the solution (level 2 prompt), showed in Figure \ref{fig:instructor-bootstrap}, contains instructions on how LLteacher should guide the student through the assignment. 
The solution can also include references to the textbooks used in the class. 

\begin{figure}[!h]
\BeginAccSupp{method=escape,ActualText={Screenshot of a simulated conversation between a student and LLTeacher in the Bootstrap Discovery assignment, showing the student being guided through the concept of bootstrap. The student also runs R code directly within their responses.}}%
\includegraphics[width=\textwidth]{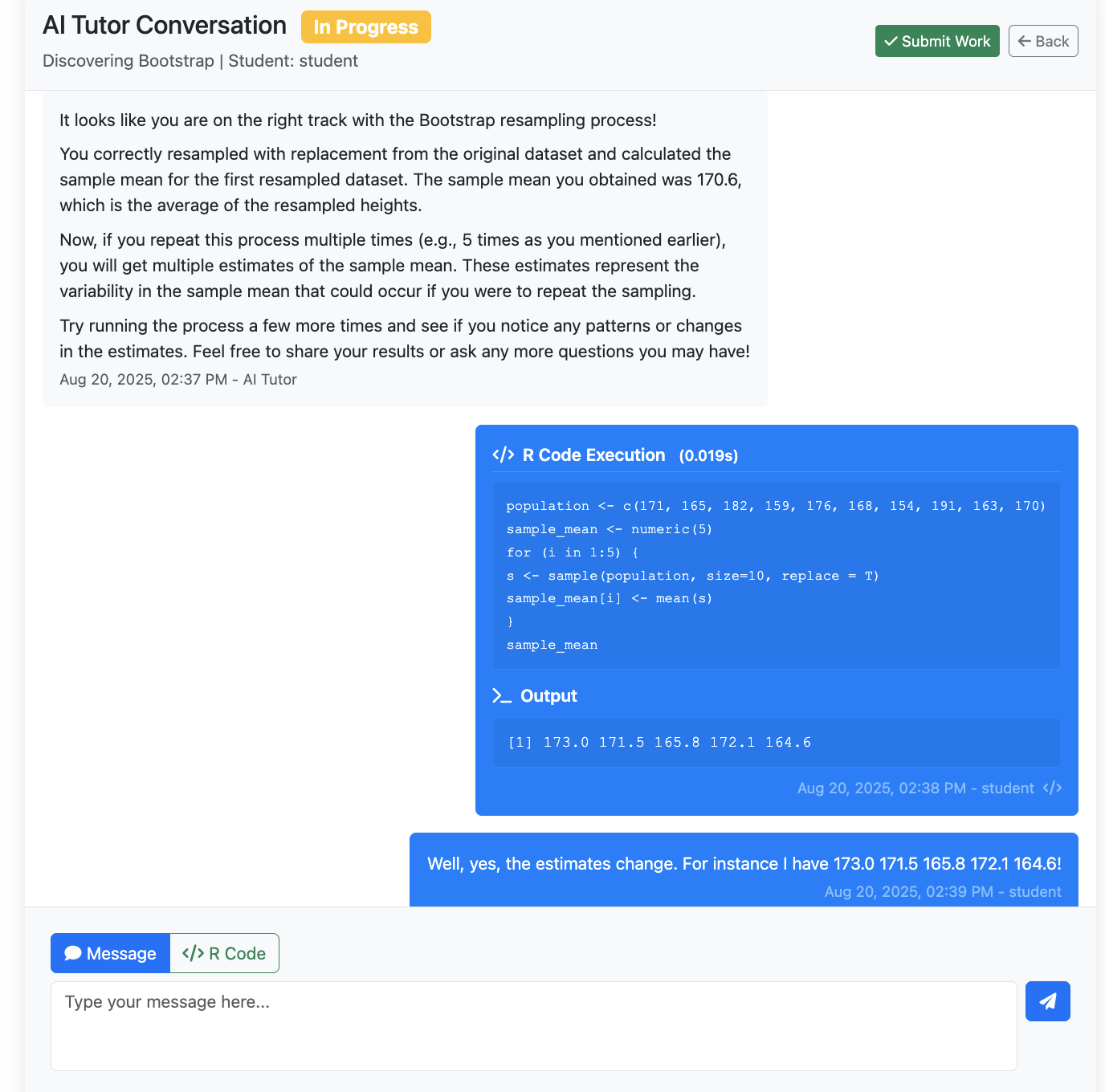}%
\EndAccSupp{}
\caption{Example of the LLteacher-student interaction in the assignment ``Bootstrap Discovery". \label{fig:student-bootstrap-2}}
\end{figure}

Figure \ref{fig:student-bootstrap-2} shows a simulated conversation between a student and LLteacher for this assignment.
This case is different from the previous one in that it does not have specific instructions for the student, who in turn has to rely a lot on the conversation with LLteacher.
The solution that the instructor writes should thus be more detailed than the one in the previous example. 
Students can also run R code directly in their answers, as shown in Figure \ref{fig:student-bootstrap-2} and as discussed in Section \ref{sec:introducing-technical-descr}.

\section{Discussion \label{sec:discussion}}

In this paper we presented the first version of LLteacher, a concrete example of how LLMs can be integrated thoughtfully into statistics homework assignments, giving instructors new tools for homework design while guiding students to engage with AI in ways that support their learning. We illustrated its usage with examples from an undergraduate statistical computing course. 

LLteacher allows instructors to provide a problem with solutions while enabling students to interact with an LLM through a sequence of interactions that guide them to the instructor-provided solution.
Unlike commonly used chatbots, such as ChatGPT, the instructor has control over the answers provided by the LLM and can view students' full interactions with the tool. 
By using LLteacher, students receive answers aligned with the course content and scope, naturally learn how to effectively interact with an LLM in a way that supports their learning, and benefit from a more equitable experience, since all students can access the same tool in the same way.
The code is fully available online making the project fully replicable by other instructors.

The design of LLteacher follows the principles that the ITS literature has identified as characteristic of effective ITS systems, and it is in line with ITS best practices: preserving productive struggle by withholding direct answers by default, matching feedback to the type of knowledge being learned via the KLI framework, and generating personalized, encouraging feedback in a human-like tone, which is a key advantage that LLM-based systems hold over the rigid, templated responses of traditional ITS \citep{stamper2024enhancing}

Beyond its connections to the ITS tradition, LLteacher also speaks to the broader debate about how to meaningfully integrate LLMs into teaching: rather than asking whether students should use LLMs, it asks \textit{how} they should.
Compared to LLM-based tutor (see for instance \citet{kazemitabaar2024codeaid}, \citet{hobert-berens-2024} and \citet{liu2025lpitutor}), LLteacher \textit{(i)} provides instructors with a new way of designing homework assignments (including discovery learning), going beyond tutoring into assignment design pedagogy; \textit{(ii)} interactions are logged for instructors to evaluate as part of the assignment; \textit{(iii)} its code is fully open-source, and \textit{(iv)} instructors can customize each assignment. It is conceived for the learning of statistics, but, thanks to its flexible design and the possibility for instructors to customize prompts, it can be adapted to other classes.


A natural concern for large-enrollment courses is whether reviewing full conversational histories is feasible at scale. LLteacher is flexible in this regard: while detailed review of student-LLM interactions enables rich qualitative feedback, the grading workflow need not depend on it. Instructors can choose to assess participation or to deploy LLteacher entirely as a non-graded support resource.

The introduction of this tool, as of other tools similar to these, in class will require instructors to create ad-hoc materials to guide students in writing effective prompts and ensure that the entire class starts at the same level. We plan on sharing instructional materials that can accompany LLteacher. These materials will specifically address how to interact constructively with an LLM: how to ask specific rather than generic questions, and how to engage in a productive back-and-forth rather than expecting a direct answer.

In the long run, we plan to further develop LLteacher to be more closely aligned with in-class materials. 
Along with homework solutions, instructors could provide, for example, their slides and notes which refer to the homework content. This would allow additional control over both the content and style of LLteacher's answers. 
The flexibility of LLteacher allows instructors to experiment with different types of assignments by modify the level 1 prompt. 
For example, instructors could instruct LLteacher to provide erroneous statements with the goal of encouraging students to critically evaluate answers from LLMs instead of blindly trusting them. 

A potential limitation of any conversational AI tool is that a student may inadvertently take the conversation off-track, for example, by asking tangential questions or pursuing an incorrect line of reasoning, causing LLteacher to follow along rather than redirect. Well-designed level 1 and level 2 prompts substantially reduce this risk, as clear instructions and a  well-defined solution space help the LLM stay focused on the assignment's  learning objectives. 
However, in the case this occurs, the instructor's ability to review the full conversation history provides a natural corrective  mechanism: instructors can identify where the interaction went astray, follow up directly with the student, and refine the prompts for future assignments.
In this way, even edge cases become opportunities for iterative improvement of the tool.

LLteacher was developed keeping in mind homework assignments in Statistics and it is still in a first version. However, its flexible framework gives it the potential to be extended to other types of assignments, such as data analysis projects, and to non-Statistics classes. 

LLM based tools have the potential to enable novel educational modalities, such as discovery-driven learning and the critical evaluation of LLM answers.
LLteacher contributes in this direction and we seek to encourage more instructors in statistics to consider using LLM based tools and sharing their experience with the community of Statistics educators.

\section{Disclosure statement}\label{disclosure-statement}

The authors have no conflicts of interest.

\section{Data availability statement}

Data sharing is not applicable to this article as no new data were created or
analyzed in this study.

\bibliography{bibliography.bib}

\end{document}